\documentclass[10pt,a4paper,twocolumn]{article}

\usepackage{lineno}
\usepackage{authblk}
\usepackage{graphicx}
\usepackage[explicit]{titlesec}
\usepackage[labelfont=bf,labelsep=endash,font=footnotesize]{caption}
\usepackage{tabu}
\usepackage{xcolor}
\usepackage[
     backend=biber, 
     natbib=true,
     style=numeric,
     maxnames=50,
     sorting=none
]{biblatex}
\usepackage[hidelinks]{hyperref}
\usepackage[hang]{footmisc}
\usepackage[normalem]{ulem}
\usepackage[top=2.5cm, bottom=2.8cm, left=1.5cm, right=1.5cm]{geometry}
\usepackage{abstract}
\usepackage{mathtools}
\usepackage{balance}
\usepackage{wrapfig}

\makeatletter
\renewcommand\AB@affilsepx{, \protect\Affilfont}
\makeatother

\providecommand{\keywords}[1]{\textbf{Keywords}\ \ \textendash\ \   #1}

\titleformat{\section}{\large\bfseries}{\thesection.}{1em}{\MakeUppercase{#1}}
\titlespacing*{\section}{0pt}{12pt}{6pt}

\titleformat{\subsection}{\large}{\thesubsection}{1em}{#1}
\titlespacing*{\subsection}{0pt}{12pt}{6pt}

\titleformat{\subsubsection}{\large\itshape}{\thesubsubsection}{1em}{#1}
\titlespacing*{\subsubsection}{0pt}{12pt}{6pt}

\newcommand{\ITUurl}[1]{\textcolor{blue}{\urlstyle{same}\url{#1}}}

\newcommand{\ITUpar}{\vspace{8pt}\par}

\renewenvironment{abstract}
               {\list{}{
               \setlength{\rightmargin}{0mm}
               \setlength{\leftmargin}{0mm}
               \vspace{-0.25in}
                \item[\textit{\textbf{\hspace{22pt}Abstract  }}  \textendash]\relax}}
               {\endlist}

\def\starttable{\vspace{6pt}\begin{table}[ht]\center}
\def\startfigure{\vspace{6pt}\begin{figure}[ht]\center}

\makeatletter
\def\tagform@#1{\maketag@@@{\ignorespaces#1\unskip\@@italiccorr}}
\makeatother

\title{\large{\textbf{x\uppercase{URLLC in 6G With Meshed RAN}}}}

\author[1]{\normalsize{Mohammad Ali Khoshkholghi}}
\author[1]{\normalsize{Toktam Mahmoodi}}
\author[2]{\normalsize{Subhankar Pal}}
\author[2]{\normalsize{Subhash Chopra}}
\author[2]{\normalsize{Mayuri Tendulkar}}
\author[3]{\normalsize{Sandip Sarkar}}

\affil[1]{\normalsize{Centre for Telecommunications Research, Department of Engineering, King’s College London, UK}}

\affil[2]{\normalsize{Capgemini, India}}
\affil[3]{\normalsize{Capgemini, USA}}

\date{\vspace{-12pt}{\small NOTE: Corresponding author: Mohammad Ali Khoshkholghi, ali.khoshkholghi@kcl.ac.uk} \\  \endgraf\rule{\textwidth}{1pt}}

\begin{document}


\twocolumn[

\begin{@twocolumnfalse}
\maketitle

\begin{abstract}
\textit{5G  Ultra-Reliable Low Latency Communications Technology (URLLC) will not be able to provide extremely reliable low latency services to the complex networks in 6G. Moreover, URLLC that began with 5G has to be refined and improved in 6G to provide xURLCC (extreme URLCC) with sub-millisecond latency, for supporting diverse mission-critical applications. This paper aims to highlight the importance of peer-to-peer mesh connectivity for services that require xURLLC. Deploying mesh connectivity among RAN nodes would add significant value to the current 5G New Radio (5G NR) enabling 6G to increase flexibility and reliability of the networks while reducing the inherent latency introduced by the core network. To provide a mesh connectivity in RAN, the nodes should be able to communicate with each other directly and be independent from the mobile core network so that data can be directly exchanged between base stations (gNBs) whereas certain aspects of signalling procedure including data session establishment will be managed by RAN itself. In this paper, we introduce several architectural choices for a mesh network topology that could potentially be crucial to a number of applications. In addition, three possible options to create mesh connectivity in RAN are provided, and their pros and cons are discussed in detail.  }
\end{abstract}

\ITUpar
\keywords{6G, mesh connectivity, RAN, xURLCC. }

\ITUpar
\ITUpar

\end{@twocolumnfalse}
]

\section{Introduction} 
Introduced in 3GPP release 15 to address the requirements of ITU-R M.2083, ultra-reliable low latency communication is, arguably, the most promising addition to the 5G capabilities. It has been further enhanced as part of 3GPP Release-16. This new URLLC wireless connectivity guarantees latency to be approximately 1 millisecond (ms) with 99.99\% reliability and is a complete game-changer for communications technology in the modern age. With it, we can conduct remote surgeries, have our cars drive for us, and increase machine productivity by large-scale factories \cite{ref12}. \ITUpar

With the advent of B5G and 6G, networks will become even more complex. There will be a network of networks, where many subnetworks roll up to a bigger network. Machine area networks such as a car area network or a body area network can have hundreds of sensors over an area of less than 100 meters. These sensors will need to communicate within 100 microseconds with extreme high reliability for the operation of that machine system \cite{ref1}. Making networks within cars or on robots truly wireless will open a new era for the designers of those devices as they would no longer need to install lengthy and bulky cable systems. Also with introduction of non-terrestrial networks, satellite and High Altitude Platform Systems (HAPS), drone networks will integrate with terrestrial networks, to create an even more complex network of networks, to provide extreme coverage also to the remotest parts of the planet. These challenges are further enhanced by elements such as expectations of node-to-node or machine-to-machine communications requirements, the uncertainty of topology, diverse application requirements, backward compatibility, user equipment resource limitations, and the rapidly increasing number of devices. These elements exacerbate the technical complications of the implementation of future B5G/6G networks. \ITUpar

5G URLCC technology will not be able to provide an extremely reliable low latency service to this complex network of networks in 6G. Moreover, URLLC that began with 5G has to be refined and improved in 6G to provide xURLCC. xURLLC promises to offer uninterrupted connectivity for a few of the interesting new services like remote surgery/patient diagnosis, live reporting of critical events like natural calamities, live sports events, metaverse applications, tactile Internet, cloud-based entertainment (VR/AR) and online gaming \cite{ref6}, \cite{ref7}. These services impose stringent QoS requirements in terms of delay ($\leq 1$ ms) with reliability (in terms of error rate) within $10^{-3}$ to $10 ^{-9}$. Both radio access technology as well as the core needs to work in tandem to ensure these high reliability and low latency requirements are met for the respective applications. One of the key components which is not completely addressed by standards pertains to communication delays and errors introduced during signalling between RAN and the core. Therefore, this needs a special focus to ensure those stringent delay and reliability requirements are met end to end. This paper proposes different options that can be included in 6G to address these aspects to ensure a seamless xURLLC user experience in 6G.  \ITUpar 

To provide xURLCC service uniformly, for such extreme network and coverage, we need an advanced network architecture that natively supports: 
i)	mesh connectivity to minimize the hops to send data between devices, and there is no need to rely on centralized core network anchor.
ii)	procedures to avoid delay due to complex signalling procedures in RAN and core networks.
\ITUpar

Hence, we aim at defining network architectures and approaches which make the mesh connectivity possible in 6G networks while taking into account the variety of applications. As our contributions, first, we introduce a particular use case considered in this research named as UE-to-UE use case with low latency communication (Section \ref{section2}). Then, given this use case, we propose several architectural designs to fulfil the diverse service demands (Section \ref{section3}). Next, we discuss the general concept of mesh connectivity as well as investigate the existing mesh connectivity provided by other wireless technologies (Section \ref{section3+1}). Furthermore, we design various architectures with a focus on the Integrated Access and Backhaul (IAB) technology where mesh connectivity is enabled among RAN nodes, although these architectures can be simply apply to regular RAN including base stations (gNBs) each of which consists of Central Units (CU) and Distributed Units (DU) (Section \ref{section4}). We also propose three solution approaches in order to create mesh connectivity in RAN as i) import core functions into RAN, ii) a peer-to-peer (P2P) RRC-based connection, and iii) a P2P XnAP-based connection, in Section \ref{section5}, Section \ref{section6}, Section \ref{section7}, respectively. To the best of our knowledge, this paper is the first work to devise a mesh-connected RAN for 6G networks in order to support xURLCC services.  \ITUpar

\section{Use Case: UE-to-UE Communication With Low Latency}
\label{section2}
\begin{figure}%
\centering
\includegraphics[width=7cm, height=8cm]{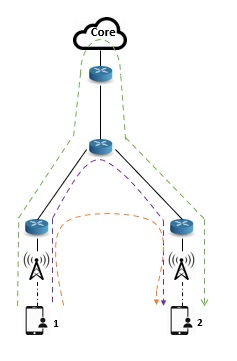}
\caption{The reference use case considered in this paper}
\label{fig0}
\end{figure}

Although significant progress has been made in reducing network latency, future applications may demand that we make much more progress in this direction. Fig. \ref{fig0} depicts the particular use case considered in this paper where different end-to-end latency targets achievable for UE-to-UE communication have been presented. The accessible latency for UE-to-UE communication relies upon the placement of User Plane Function (UPF) anchoring the Protocol Data Unit (PDU) sessions of UEs. Consider the scenario in Fig. 1 where UPF is located in the remote core network, then data should pass through the core network (green line) even if the UEs are connected to two neighbouring gNBS. This may increase the end-to-end latency for the UEs' communications to tens of ms. In the other case, UPF could be placed in the aggregation site closer to UEs forming the purple data path where the end-to-end latency can be decreased up to few ms. Finally, by deploying UPF in the RAN nodes, a direct communication between gNBs (orange data path) allows even lower latency targets required for the new applications of 6G networks. This scenario relies on deploying a mesh connectivity among RAN nodes which is our main objective in this work. \ITUpar

 It is worth mentioning that Dual Connectivity (DC) and Side Link (SL) communications (via PC5 interface) are other options which could be considered for UE-to-UE communication. In DC, each UE is connected to two base stations simultaneously whereas SL deploys a direct communication between two UEs without the participation of a base station in data transmission \cite{ref5}. However, since the main scope of this research is to study the possibility of establishing mesh connectivity in NR RAN (among base stations), we leave DC and SL communications for our future works. \ITUpar

Each of these scenarios can be more appropriate for a sort of use cases and applications with particular requirements, so that we can design different architectures to support the corresponding services. In addition, deployment of a multi-connectivity scheme derived from a mesh-based network architecture creates multiple redundant paths between UEs enabling reliable and stable communications.\ITUpar

\section{Consideration of Network Architecture}\label{section3}

\begin{figure*}[t]

    \includegraphics[width=18cm]{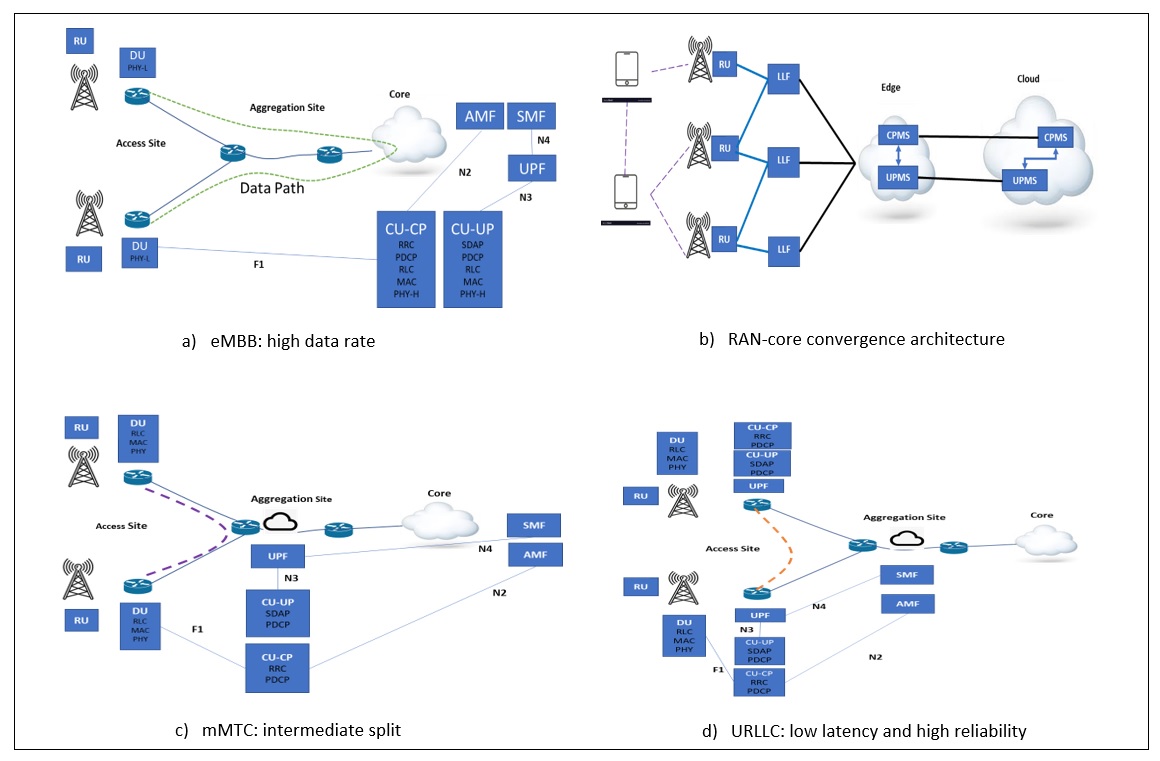}
    \caption{Different architectural designs for the introduced use case. }
    \label{fig1}
\end{figure*}

In this section we present four architecture formats, which are designed for different 6G use cases with various requirements in terms of latency, reliability, data rate and RAN node complexity; this diversity is seen in Fig. \ref{fig1}. Generally, the enhanced mobile broadband (eMBB) services need high bandwidths, that are not necessarily latency-sensitive. Hence, a high degree of centralization for tight coordination across gNBs should be provided to achieve high data rates. Fig.  \ref{fig1}(a) shows the appropriate architecture for these types of services where the core network functions (e.g. Access and Mobility management Function (AMF), Session Management Function (SMF) and User Plane Function (UPF)) are co-located with RAN Central Unit User Plane (CU-UP) and Central Unit Control Plane (CU-CP) functions. We consider RAN functional split option 7 to centralize most of the functions in the core network. In option 7 \cite{ref8}, all the higher layers of both the CU-CP and CU-UP are located in the core, while the access site including distributed unit and radio unit only host lower functionalities of the physical layer (PHY-L) and Radio Frequency (RF) function, respectively. DUs are kept as light components decreasing the RAN node complexity in this architecture. Light components refer to the functions where most or some of their functionalities are shifted to the core or CU. Since most of the functions are centralized in the core in this architecture, the data path as well as signalling messages for the session management pass through the core (green line in the figure).  

We can extend the architecture proposed for the eMBB case further to the architecture depicted in Fig. \ref{fig1}(b) leveraging RAN-core convergence which is one of the objectives defined for 6G networks \cite{ref2}. In this architecture, core functions and the majority of RAN functionalities are moved to a central cloud and developed as cloud-native micro-services (Control Plane Micro-services (CPMS) and User Plane Micro-services (UPMS)). In this architecture, each RAN node consists of an RU as well as the Lower Layer Function (LLF) as a light entity including all the air-interface-related RAN functionalities that are not included in the RU. These micro-services could be shifted from the remote cloud to an edge node closer to the users as well.  \ITUpar

Fig. \ref{fig1}(c) shows another architecture proposed for the use cases where latency matters but is not vital such as mMTC services. To this end, although the control plane core functions (AMF and SMF) are kept in the core, the UPF is co-located with RAN CU functions in the aggregation site considering an intermediate RAN functional split option (option 2 or 3). In this case, although the signalling messages still go through the core, data passes through aggregation site with lower latency than the previous architectures (purple line in the figure). \ITUpar

The last proposed architecture is shown in Fig. \ref{fig1}(d) for the Ultra-Reliable and Low Latency Communication (URLLC) use cases. The URLLC users require a decentralized split, which minimizes the latency experienced by these services. Hence, most functions are decentralized and situated in the access site. Low latency and high reliability could be ensured by the increased degree of decentralization. The core control plane functions are located in an aggregation site (e.g. edge node) while UPF and RAN functions are decentralized in the access network. This scenario is based on deploying a mesh connectivity among the base stations upon which gNBs are able to send data to each other directly (orange path in the figure). However, signalling messages still pass through the core functions located in the aggregation site. \ITUpar

\section{Mesh Connectivity in other wireless communication Technologies } \label{section3+1}

In this section, we explain the general concept of mesh connectivity and explore the existing mesh connectivity provided by other communication technologies including Bluetooth and WiFi. A mesh network is an interconnected communication network made up of different devices and nodes (physical redistribution points which receive and transmit wireless signals). Because of their decentralized nature, mesh networks can continue to scale almost endlessly, maintaining signal strength and the ability to send and receive data with a high degree of reliability. As the number of connected devices scales rapidly (a predicted growth of almost 5x in a 10-year span), mesh networks will enable consumers and businesses to connect all their devices without the need for dedicated hubs. This allows for the proliferation of networks of connected things. Mesh networks are self-healing, meaning that if there is any disruption to the connectivity of a certain device, the network just connects to other devices and the network is not dropped. \ITUpar

Mesh networks are very popular in Wi-Fi and Bluetooth. Bluetooth mesh networking \cite{ref13} is a new topology available for Bluetooth Low Energy (LE) devices that enables many-to-many communications. It's optimized for creating large-scale node networks and is ideally suited for building automation, sensor networks, and asset tracking solutions. It is a reliable method of sharing information in large networks. Bluetooth mesh operates under one principle: a flood network. This is where the nodes relay messages in a certain manner, either in uncontrolled flooding or controlled flooding. Bluetooth mesh has a message cache that prevents conveying similar messages or messages relayed recently. It’s more reliable and easier to install, and enables vendor interoperability,  range extension and power saving. 

In the context of WiFi, a Wireless Mesh Network (WMN) \cite{ref14} is a mesh network created through the connection of Wireless Access Point (WAP) nodes installed at each network user's locale. It often consists of mesh clients, mesh routers and gateways. The networking infrastructure is decentralized and simplified because each node needs only to transmit data as far as the next node. The network topology of a wireless mesh network may be full or partial mesh. A full mesh network means every node communicates with every other node. In a partial mesh topology, nodes only communicate with nearby nodes. When data is transmitted between two nodes that do not communicate with each other, data hops from one node to the next until it reaches the destination. The nodes are programmed to use adaptive routing algorithms to constantly determine the optimal route between nodes for data transmission. This networking architecture provides collaborative, redundant backup technology, which ensures data security in the event of disk failure. It offers increased reliability, as each node is connected to several other nodes and, if one drops out of the network, its neighbours simply find another route and its scalable.

\ITUpar

\section{IAB architectural design } \label{section4}

\begin{figure*} [t]

    \includegraphics[width=18cm]{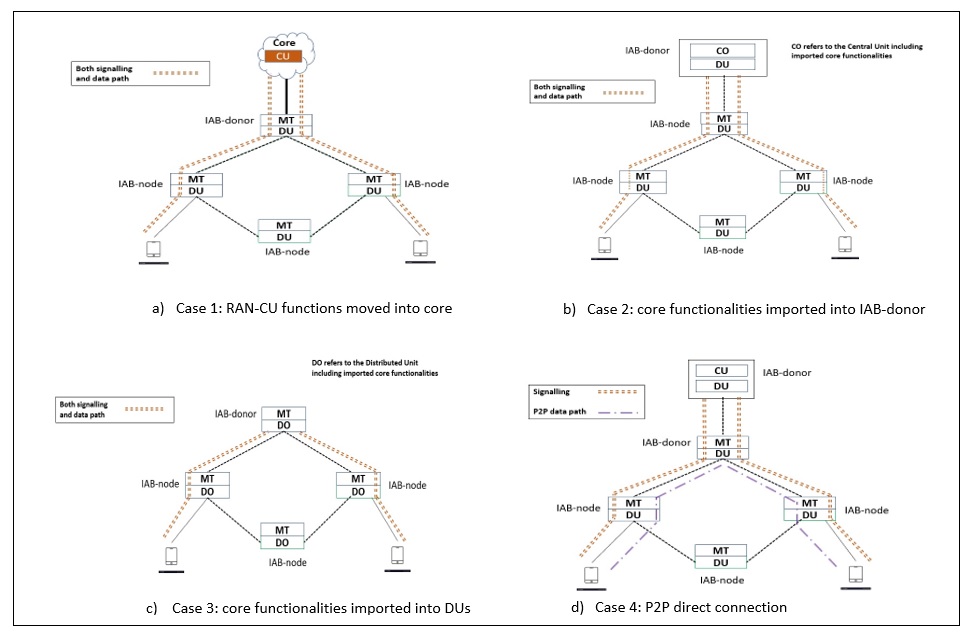}
    \caption{Different architectural designs for IAB. }
    \label{fig2}
\end{figure*}

With 5G millimeter Wave (mmWave) bringing very high speeds and capacities, it also brings the limitation due to its limited coverage at high frequencies. So, the carriers need many more cells to reach denser areas. To overcome this challenge and facilitate faster 5G deployments, 3GPP has standardized a solution for multi-hop relaying support over 5G NR in Release 16 called Integrated Access and Backhaul (IAB) that is of particular interest for dense deployment of street-level radio nodes. IAB architecture \cite{ref3} allows for multi-hop backhauling and includes an IAB-donor node which is wire-connected to the core network and hosts RAN-CU functionalities, along with IAB nodes connected to IAB-donor node and to each other wirelessly. Each IAB-node consists of two entities as: i) DU which connects the node to both end users and downlink IAB nodes, and ii)  Mobile Terminal (MT) connecting the node to uplink IAB-nodes or an IAB-donor node. The structure and functionalities of entities in the IAB architecture is similar to WiFi. However, mesh connectivity between gNBs has not yet been explored in the IAB architecture, unlike WiFi. Therefore, considering the IAB architecture, in this section we try to find an appropriate approach that enables mesh connectivity. \ITUpar

As discussed in Section \ref{section3}, based on the placement of the UPF, there might be different path options for data forwarding between base stations, each could be efficient for a sort of services. However, to enable the shortest path communication in order to achieve tight latency requirements of 6G use cases, a direct connection between gNBs is necessary to create a mesh connectivity in RAN. In the last architecture introduced in Fig. \ref{fig1}(d), although data can be sent directly between gNBS, signalling messages still need certain core network procedures to manage the session establishment. However, in order to establish a mesh connectivity in RAN, we aim at designing architectures which handle the direct connection between gNBs without the need for the core functions. In this section, we propose four different IAB architectures designed with/without core functions as shown in Fig. \ref{fig2}.  It is worth mentioning that we consider scenarios that an entire call can be contained inside RAN or a group of RANs cooperating with each other. In other words, our work considers coreless RAN communication. \ITUpar

Fig. \ref{fig2}(a) shows the first proposed architecture where RAN-CU functions are moved from an IAB-donor to the core network. This architecture is in line with the first and second architectural options introduced in Section \ref{section3}, where most RAN functionalities are co-located in a core/cloud while keeping RAN nodes as  light entities. In this architecture, both data and signalling messages go through the core/cloud to serve use cases requiring a high data rate. \ITUpar

The second architecture is depicted in Fig. \ref{fig2}(b). In this option, we import and adapt all the functionalities of a core network into a RAN-CU located in IAB-donors (or gNBs) so that each RAN node will be able to connect to another node independently and without a need to traverse the core network. We will discuss the details of this approach in Section \ref{section5}. Hence, in this architecture neither data nor signalling messages go through the core in case of a direct communication between nodes. Using this architecture, two IAB-nodes are able to send data directly to each other while signalling messages pass through a CU located in the IAB-donor. DUs are light entities in this case.\ITUpar

In the third architecture (Fig. \ref{fig2}(c)) we import the core functionalities directly into the DUs of IAB-nodes so that they can create a mesh connectivity among themselves even without passing an IAB-donor for signalling. It can provide fast connection between IAB nodes and increase the flexibility of IAB architecture, however, DUs will not be light entities anymore since we should centralize RAN and core functionalities in DUs which consequently increases the node complexity and deployment cost.   \ITUpar

Although we are able to remove core dependency for creating mesh connectivity using the second and third architecture, but we need a considerable amount of modifications on the current 5G NR technology and standards since we should develop all the required core functions in the RAN nodes. To avoid this, in the fourth architecture we present a solution upon which we can provide mesh connectivity among RAN nodes (gNBs or IAB-donors/nodes) without importing core functionalities into RAN. As seen in Fig. \ref{fig2}(d), a CU including usual RAN-CU functions is located in the IAB-donor, but there will be direct P2P connections among RAN nodes (e.g. IAB-nodes) without interference of core functions. We will discuss this solution in detail in Section \ref{section6} and Section \ref{section7}.   \ITUpar

\section{Development of core functionalities inside RAN nodes} \label{section5}

To deploy mesh connectivity in RAN where nodes are able to communicate with each other directly without the need for core functions, one approach is to centralize all the connectivity features within RAN nodes. The core functionalities can be added to and/or converged with RAN functions co-located with RAN CU or DU entities. Given the standard user plane and control plane protocol stacks of RAN nodes \cite{ref3}, one protocol layer named the mesh layer should be adopted and developed on top of the user plane and control plane so that it covers all the required connectivity features previously supplied by the core network in addition to mesh-related features. This layer comprises several components as follows: 
\begin{itemize}
  \item \textbf{Access and handover management:} this component is implemented on top of the RRC layer in the control plane protocol stack and is in charge of access authentication and authorization of UEs, policy enforcement, security context management and handover management of UEs. 
  
  \item  \textbf{Resource scheduling and QoS management:} this component schedules and manages the radio resources of RAN nodes while assuring the QoS required by different services. This layer is developed on top of the XnAP layer in the control plane protocol stack and the session establishment between RAN nodes will be handled and managed by this component. 
  
  \item \textbf{Dynamic network management:} deployed on top of XnAP, this component organizes and manages the mesh connectivity-related features in a self-organized RAN.
  
  \item \textbf{Data forwarding:} this component supports packet routing and forwarding towards RAN nodes across the network, and is developed on top of the SDAP layer in the user plane protocol stack.  \ITUpar   
  
  \end{itemize}
  
  This approach is one of the most efficient in terms of reduction in delays and also error rates by collocating key core functions alongside CU/DU and IAB donors. In addition, this approach can be suitable for the use cases which require an adaptive RAN where connecting to a remote core or backhaul node is difficult or costly like rural or remote areas (e.g a maritime use case). On the contrary this approach introduces complexity in terms of introducing new components co-located with CU/DU which are currently not part of 3GPP SA work items. \ITUpar

\section{P2P RRC-based Connection} \label{section6}

\begin{figure}%

    \includegraphics[width=8.5cm]{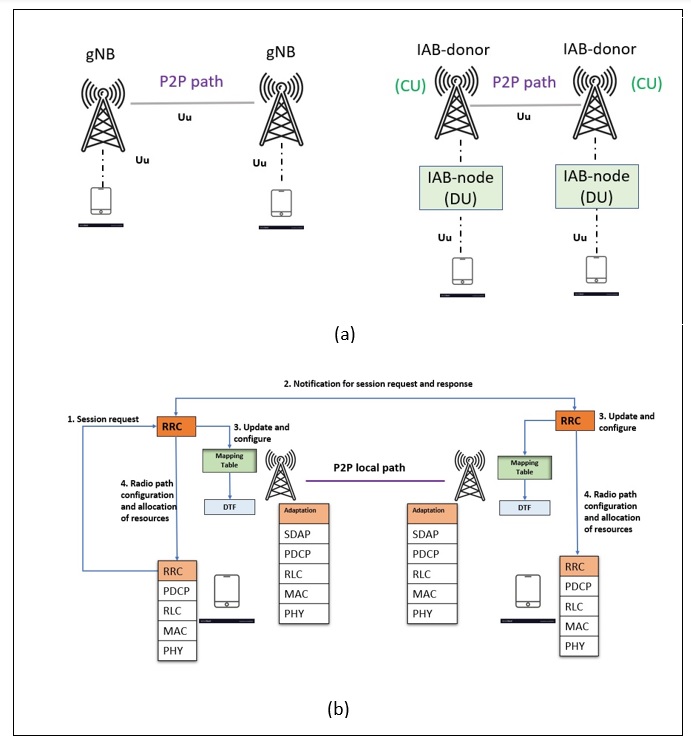}
    \caption{P2P RRC-based connection: a) direct connection between base stations or IAB donors; b) direct connection procedure and protocol stacks}
    \label{fig5}
\end{figure}

\begin{figure*} [t]
     
    \centering
    \includegraphics[width=14cm]{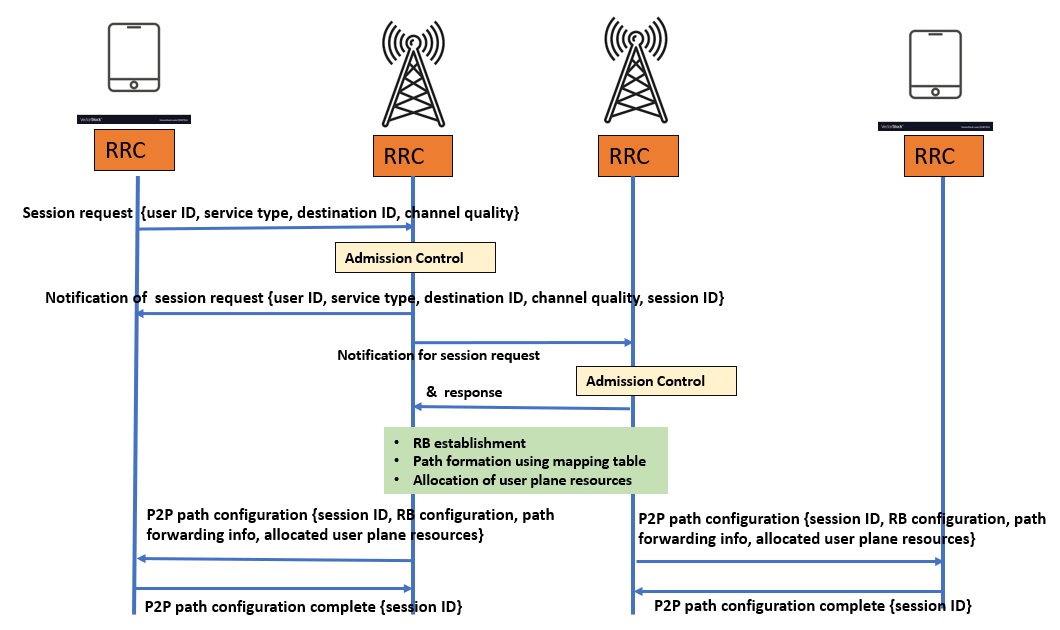}
    \caption{P2P RRC-based connection: signalling}
    \label{fig6}
\end{figure*}

\ITUpar 

As discussed earlier, the solution introduced in Section \ref{section5} is able to create a coreless mesh-connected RAN. However, in order to deploy this solution, the current 5G NR standards need to be modified significantly. Therefore, another solution which provides the same benefits with less modifications is introduced by which we can take advantage of mesh connectivity in less time and deployment cost. In this section, we introduce this approach which extends a vehicular communication approach \cite{ref15} used to transmit localised V2X traffic, into the context of IAB architecture. This approach is able to create direct communication between RAN nodes (gNBs or IAB-donors) without core network interference using an RRC-based connection as shown in Fig. \ref{fig5}. In this approach, we install a P2P data path between a pair of nodes to send data through air interface ($U_u$) (Fig. \ref{fig5}(a)), whereas signalling procedure happens through the RRC layer on top of the control plane protocol stack in order to establish and manage PDU sessions between pairs. \ITUpar

The procedure of this approach is summarized in Fig. \ref{fig5}(b). Consider a UE (left one in the figure) aims to communicate with another UE via a direct connection between gNBs. In this case, a session request from the RRC of the UE is sent to the RRC of the connected gNB. Then, the RRC of the gNB sends a notification message to the corresponding gNB and gets the response back from it. The RRC, then, is able to update and configure a mapping table based on the information received from other gNBs by which the gNB will direct and transmit the UE's data to the destination. Since this approach works independently from the core functions, a Data Transfer Function (DTF) is developed to perform the functionalities similar to the UPF in order to transmit data through a P2P path, such as application detection, packet routing and forwarding, per-flow QoS handling, and traffic usage reporting. As can be seen, two new components (DTF and mapping table) are adopted as a sublayer on top of the user plane protocol stack which helps to transmit data directly between gNBs without the need to use the core network. Fig. \ref{fig6} shows the signalling procedure of the P2P RRC-based connection between UEs and gNBs. Once a UE aims to connect to another UE, it sends an RRC session request to the source gNB including user ID, service type, destination ID and channel quality. The service type information provides the QoS requirements to the gNB such as latency and reliability requirements. The channel quality information provides the required data plane resources to be allocated for the session being established. The RRC of the source gNB checks the available user plane resources and sends back a response including the session ID which is the identifier of radio bearers allocated to the given service request.\ITUpar

In the next step, the gNB sends a notification for the session request to the target gNB where its RRC checks the possibility of creating a connection according to the available resources as well as QoS requirements for the given service request and sends back the response. Next, the source and target gNBs perform several tasks including radio bearer establishment, P2P path creation between UEs through gNBs using the information derived from the mapping table, allocating the user plane resources for the session, as well as configuring and updating the mapping table. In the next step, source and target gNBs send a path configuration message to the corresponding UEs. The path configuration message includes session ID, radio bearer configuration, path forwarding information and allocated user plane resources. Radio bearer configuration provides required information for establishment of radio bearers and path forwarding information includes mapping table information to transmit the data over the P2P path. In the last step, UEs send back a completion message confirming the session establishment. Finally, the data flow will be transferred over the P2P path.  \ITUpar

Since this approach uses air interface ($u_u$) to provide mesh connectivity among RAN nodes, it is more suitable for IAB technology and V2X communication applications where nodes are connected through ($u_u$).   \ITUpar

\section{P2P XnAP-based Connection}\label{section7}

\begin{figure} [t]
     
    \centering
    \includegraphics[width=9cm]{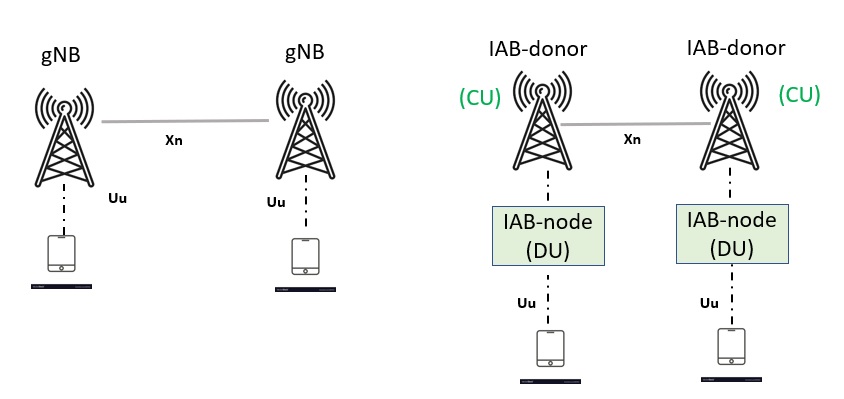}
    \caption{ Direct connection between base stations or IAB donors via XnAP }
    \label{fig5a}
\end{figure}

\begin{figure*} [t]
     
    \centering
    \includegraphics[width=14cm]{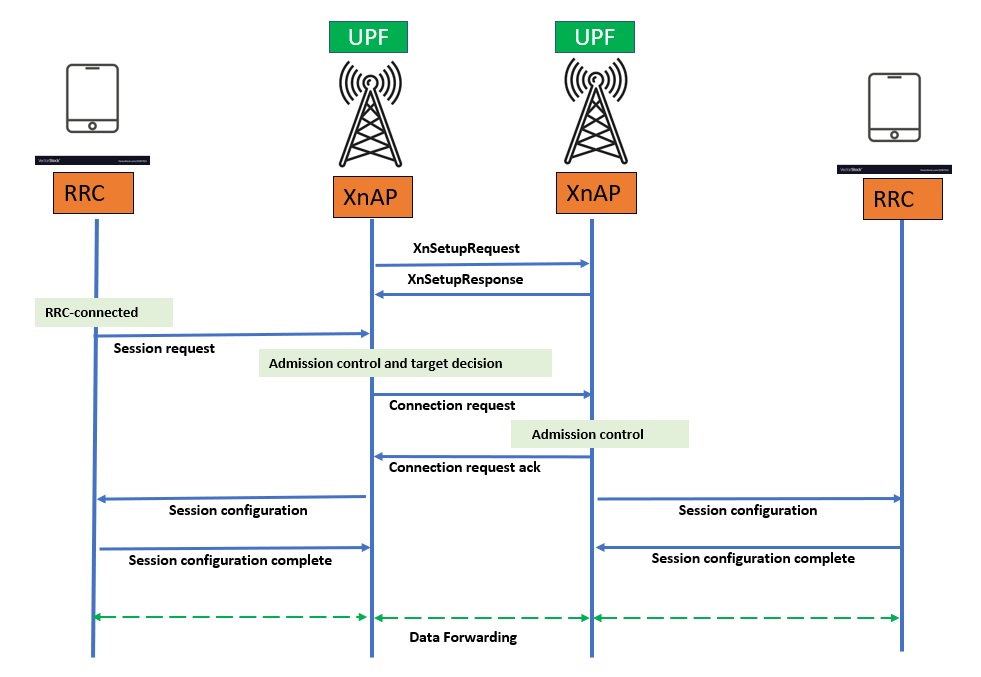}
    \caption{ Signalling procedure of the Xn-based connection}
    \label{fig5b}
\end{figure*}

\ITUpar

In the previous section, we explained the RRC-based connection in order to create direct communication between RAN nodes without the need for core functions. Although this approach needs less modifications on the current 5G NR technology compared with the first approach introduced in Section \ref{section5}, still we need to develop a sublayer on top of the user plane protocol stack as well as to modify the current $U_u$ interface according to the signaling procedure between gNBs discussed earlier. However, as mentioned before, in order to accelerate the deployment process and reduce the deployment cost upon the existing 5G technology, in this section we propose another solution which tries to create mesh connectivity with the least modifications. \ITUpar

In this approach, as shown in Fig. \ref{fig5a}, unlike the RRC-based connection which establishes a P2P path through air interface ($U_u$) between RAN nodes, the direct connection will happen through the regular and standard interface connecting gNBs ($X_n$ interface) in 5G NR\cite{ref10}. This approach uses Xn application layer ($X_nAP$) \cite{ref9} of the $X_n$ interface within gNBs to connect them to each other so that data and signalling messages forward through the $X_n$ interface without traversing the core network. However, to be able to send data directly between gNBs, we consider a UPF instance is located in each gNB enabling data forwarding over the Xn. At the current 5G network architectures, UPFs are located on a central cloud or possibly on edge nodes closer to the RAN. However, in this approach UPF instances are shifted towards the RAN in order to provide a direct, fast and reliable connection between gNBs. Therefore, we do not need to add any sublayer to the existing user plane protocol stack anymore. \ITUpar 

In the current 5G NR technology, the Xn interface has been used for two purposes: a) user handover between gNBs, and b) in the Non-Standalone (NSA) architecture to connect gNBs and 4G eNBs. However, in this approach, we adapt the Xn interface in a way that it can be used for direct communication between RAN nodes and consequently to enable mesh connectivity in the RAN. For this, we consider the user handover procedure currently deployed in 5G NR \cite{ref11}, when a UE moves from one gNB to another; and extend and modify it so that it can be applied for establishing direct gNB connections. The signalling procedure of the proposed $X_nAP$-based connection is presented in Fig. \ref{fig5b} modifying the existing handover procedure. In the initial step, XnAP modules in the source and target gNBs create an active connection using XnSetupRequest and XnSetupResponse messages, respectively. In addition, the UE is in the RRC-connected state to send and receive uplink and downlink data to/from the source gNB. In the next step, the UE aiming to connect another UE through a direct connection between gNBs, sends an RRC session request to XnAP of the source gNB over the Xn interface. This session request includes user ID, service type, destination ID, and the list of requested PDU sessions in the form of a transparent RRC container. \ITUpar

The XnAP module in the source gNB checks whether there are available resources to admit the session request and if so, it determines which target gNB should be connected with for the given request, and then sends a connection request to the XnAP module of the target gNB. In the next step, the target gNB performs the admission control and if it decides to admit the request, allocates the resources to the UE request and sends an  RRC acknowledge message to the source gNB. The acknowledge message is transferred as a transparent RRC container including the list of admitted and not admitted PDU sessions. Next, the source and target gNBs send the RRC session configuration messages to the corresponding users containing the required information for the PDU session establishment including session ID, bearer information, path forwarding information and allocated user plane resources. Finally, UEs send a completion message confirming the session establishment to the gNBs. After successful establishment of the session and connection of the UEs, data is forwarded over the Xn interface using UPF instances located in RAN nodes.  \ITUpar

This approach uses the Xn interface (the existing interface between gNBs) to provide mesh connectivity within the RAN. Hence, it could be more suitable for general, and a wide range of 6G applications, where gNBs connecting to each other through the Xn will create a mesh-enabled RAN. Table 1 summarizes and compares three proposed solutions and outlines the benefits and drawbacks or considerations of each solution.

\begin{table*}  [hbt!]
\centering
\caption{{Comparison of three approaches proposed in this work}}\label{tab:tab1} 
\begin{small}
\begin{tabu}{|p{2.6cm}|p{1.4cm}|p{3.9cm}|p{3.9cm}|p{3.9cm}|}
\hline
 \rowfont{\bfseries}Approach & Interface \newline & Application & Benefits & Considerations   
  \\
\hline
\hline
Developing core \newline functionalities inside RAN nodes \newline & $X_n$ & For the use cases which require an adaptive RAN where connecting to remote core or backhaul node is difficult or costly like  rural or remote areas (e.g maritime use case) \newline & Most efficient in terms of
reduction in delays and error rates as well as providing self-organized mesh-connected RAN & Introduces complexity in terms
of introducing new components co-located with CU/DU which are currently not part of 3GPP SA work items  

 \\
\hline
 P2P RRC-based Connection  \newline & $U_u$ & More suitable for IAB technology and V2X communication applications where nodes are connected through $U_u$ \newline & Providing direct wireless connection between RAN nodes using air interface  & Needs to develop a sublayer on top of user plane protocol stack as well as to modify the current $U_u$ interface  \\
 \hline
 P2P XnAP-based connection \newline & $X_n$ & For general and a wide range of 6G applications where gNBs connect to  each other through $X_n$ \newline  & Providing mesh connectivity in RAN with the current 5G NR technologies  & Needs to modify the current $X_n$ interface    \\
\hline

\end{tabu}
\end{small}
\end{table*}
\ITUpar

\section{Conclusions and Future Work}

Recognizing its importance to facilitate faster 5G deployments, in this paper, we aimed to enable mesh connectivity among RAN nodes. We proposed several architectural choices where mesh connectivity can be supported in 6G networks. In order to create a mesh connection between gNBs, they should communicate with each other in a way that data can be transferred directly between gNBs, whereas the signalling procedure should be handled within the RAN, without passing through the core network. To this end, we introduce three approaches to manage PDU session management inside the RAN, as well as providing the capability of direct data forwarding among gNBs without core network interference. In the first approach, the core functionalities should be imported and adapted within the RAN so that data and signalling messages can be sent directly between gNBs. In the second approach, an RRC-based connection is used to make a direct communication between gNBs and then data will be sent through a P2P local path. Finally, an XnAP-based connection approach is proposed where it makes possible direct connection among RAN nodes with the least amount of modifications on the current 5G NR technology. For future work we aim to implement our work on a real-world testbed to evaluate the efficiency of the proposed solutions compared with the current URLLC services provided by 5G networks. In addition, we will deploy mesh connectivity considering sidelinks connecting UEs as well, in order to support other use cases that will emerge in future mobile networks.  \ITUpar


\printbibliography 

\section*{Authors}
\label{sec:auth}


\textbf{Mohammad Ali Khoshkholghi} received his Ph.D. degree in computer science from University Putra Malaysia in 2017, and bachelor's and master's of computer science, in 2007 and 2011, respectively. He is currently a research associate at the Centre for Telecommunication Research (CTR), Department of Engineering, King's College London (KCL), UK. Before joining KCL, he worked as a postdoctoral research fellow with the DISCO Research Group, Department of Computer science, Karlstad University, Sweden, from 2018 to 2020. He has also worked as a researcher and university lecturer within computer science in industry and academia. He serves as a referee, TPC and editorial board member for many prestigious journals and conferences. His research interests lie in the area of edge and cloud computing, 6G networks and machine learning.  \ITUpar

\textbf{Toktam Mahmoodi} is a professor of communication engineering and Director of Centre for Telecommunications Research (CTR) at King's College London. She received a B.Sc. degree in electrical engineering from the Sharif University of Technology, Iran, and a Ph.D. degree in telecommunications from King’s College London, UK, in 2002 and 2009 respectively. She was a visiting research scientist with F5 Networks, San Jose, CA, USA, in 2013, a postdoctoral research associate with the ISN Research Group, Electrical and Electronic Engineering Department, Imperial College, from 2010 to 2011, and a mobile VCE researcher, from 2006 to 2009. She has also worked in the mobile and personal communications industry, from 2002 to 2006. She has contributed to, and led a number of FP7, H2020, and EPSRC funded projects, advancing mobile and wireless communication networks. Her research interests include 6G mobile networks, open networking, critical communication and and low latency networking.\ITUpar

\textbf{Subhankar Pal}  is Senior Director and global Innovation leader for the Intelligent Networks program in Capgemini Engineering. He has close to 22 years of experience in the telecommunication industry. His responsibilities include technology product incubation, product strategy and roadmap development, and consulting services offering definition for telecommunication and related markets. His interest areas include advanced network automation, optimization and sustainability using cloud native principles and machine learning for 5G and beyond networks. Prior to joining Capgemini, he worked at Nokia Networks and C-DOT. Subhankar has extensive experience in speaking at international conferences and presenting technical papers in various forums. He has several blogs and whitepapers published in international journals and event proceedings.
 
 \ITUpar

\textbf{Subhash Chopra}  has 23 years of professional experience in the product engineering services and telecommunications industry with a focus on technology and innovation. He has extensive experience in solution architecture development, system integration and technical consulting for complex telecommunication networks and related services. 
Subhash is associated with Altran, a Capgemini Group company, since 2001. He is part of the Research and Innovation Team with a focus on “5G ORAN RIC” and “Industry 4.0” solutions. His responsibilities include incubating new solutions, research on new areas related to B5G/6G, developing proof-of-concepts, developing roadmaps for productization of solutions, and services offering definitions and technology consulting.

 \ITUpar

\textbf{Mayuri Tendulkar} received a B.E. degree in electronics engineering from Mumbai University, India and M.S. degree in electrical engineering, telecommunications from the University of Maryland Baltimore County, USA. She is currently part of the Capgemini Global Innovation Team Connectivity Group and working on intelligent network solutions with 5G O-RAN RIC and NWDAF. Her  research  interests  include 5G and beyond, network virtualization, network sustainability, intelligent networking and embedded platforms. \ITUpar 

\textbf{Sandip Sarkar}  graduated from the Indian Institute of Technology in Kanpur, India with a bachelor’s degree in technology. He earned his Ph.D. in electrical engineering from Princeton University, and is currently the lead of 5G strategy for Capgemini. Before that Sandip was a senior director at Qualcomm the worldwide lead for 5G/LTE. Sarkar was initially hired by Qualcomm as a senior engineer and worked on Globalstar modem design and Condor encryption algorithm design. Later, he became the cdma2000 reverse link systems lead, and subsequently served as the editor for the cdma2000 physical layer and band class specifications. Then, he served as the MIMO lead for 3GPP, the UMB and LTE standards delegate for Qualcomm, and the editor for UMB physical layer specifications as well.
He has served as a chairman of the physical layer text group as well as an editor for 3GPP2 specifications. Sarkar is a senior member of IEEE, and holds more that 300+ worldwide patents.\ITUpar

\end{document}